\documentclass[authoryear,preprint,a4paper,12pt]{elsarticle}

\usepackage{url}
\usepackage{multirow}
\usepackage{subfig}
\usepackage{rotating}
\usepackage{graphicx}
\usepackage[english]{babel}
\usepackage{amsfonts}
\usepackage[usenames]{color}

\journal{} 

\begin{document}
\begin{frontmatter}
   \title{Design, Engineering, and Experimental Analysis of a Simulated Annealing
     Approach to the Post-Enrolment Course Timetabling Problem}
  \author{Sara Ceschia, Luca Di Gaspero, Andrea Schaerf}
  \address{DIEGM, University of Udine\\
    via delle Scienze 208, I-33100, Udine, Italy}
  \ead{\{sara.ceschia,l.digaspero,schaerf\}@uniud.it}

  \begin{abstract}
The post-enrolment course timetabling (PE-CTT) is one of the most studied timetabling problems, for which many instances and results are available. In this work we design a metaheuristic approach based on Simulated Annealing to solve the PE-CTT. We consider all the different variants of the problem that have been proposed in the literature and we perform a comprehensive experimental analysis on all the public instances available. The outcome is that our solver, properly engineered and tuned, performs very well on all cases, providing the new best known results on many instances and state-of-the-art values for the others.
  \end{abstract}


\end{frontmatter}

\section{Introduction}

Timetabling problems are widespread in many human activities and their
solution is a hard optimisation task that can be profitably tackled by
Operations Research methods. Educational timetabling is a sub-field of timetabling that
considers the scheduling of meetings between teachers and students.

A large number of variants of educational timetabling problems
have been proposed in the literature, which differ from each other
based on the type of institution involved (university, school, or
other), the type of meeting (course lectures, exams, seminars, \dots),
and the constraints imposed.

The \emph{university course timetabling} (CTT) problem is one of the
most studied educational timetabling problems and consists in
scheduling a sequence of events or lectures of university courses in a prefixed
period of time (typically a week), satisfying a set of various
constraints on rooms and students. Many formulations have been
proposed for the CTT problem over the years. Indeed, it is
impossible to write a single problem formulation that suits all cases
since every institution has its own rules, features, costs, and 
fixations.

Nevertheless, two formulations have recently received more attention
than others, mainly thanks to the two timetabling competitions, ITC
2002 and ITC 2007 \citep{MSPM10}, which have used them as competition
ground.  These are the so-called \emph{curriculum-based course
  timetabling} (CB-CTT) and \emph{post-enrolment course timetabling}
(PE-CTT). The main difference between the two formulations is that in
the CB-CTT all constraints and objectives are related to the concept
of \emph{curriculum}, which is a set of courses that form the
complete workload for a set of students. On the contrary, in PE-CTT
this concept is absent and the constraints and objectives are based on
the student enrolments to the courses.

In this work we focus on the PE-CTT problem and we design a
single-step metaheuristic approach based on Simulated Annealing (SA),
working on a composite neighbourhood composed of moves that reschedule
one event or swap two events. The solver is able to deal with all the
different variants of the PE-CTT problem proposed in the literature.

We experiment our solver on all the instances that have been made
publicly available (up to our knowledge).  The outcome of our
experimental analysis is that our general solver, properly engineered
and tuned, is able to outperform most of the solvers specifically
designed and tuned for a single specific formulation and/or a specific
set of instance.

\section{Problem Definition}

Over the years, different versions of the PE-CTT problem have been
defined. We first illustrate (Section~\ref{sec:definition}) the most
general version, which is the one that has been used for ITC 2007
and is described by \citet{LePM07}.
The other versions are obtained from this one by removing some of the
features, and they are described in Section~\ref{sec:instances} along
with a presentation of the available instances.

\subsection{General Definition of PE-CTT}
\label{sec:definition}

In the PE-CTT problem it is given a set $\mathcal{E} = \{ 1, \dots, E\}$ of events, a set
$\mathcal{T} = \{ 1, \dots, T\}$ of timeslots, and a set $\mathcal{R} = \{ 1, \dots, R\}$ of rooms. It is
also defined a set of days $\mathcal{D} = \{ 1, \dots, D\}$, such that each timeslot
belongs to one day and each day is composed by
$T/D$ timeslots.

It is also given a set of students $\mathcal{S}$ and an enrolment
relation $\mathcal{M} \subseteq \mathcal{E} \times \mathcal{S}$, such
that $(e,s)\in \mathcal{M}$ if student $s$ attends event $e$.

Furthermore, it is given a set of features $\mathcal{F}$ that may be
available in rooms and are required by events. More precisely, we are
given two relations $\Phi_R \subseteq \mathcal{R} \times \mathcal{F}$
and $\Phi_E \subseteq \mathcal{E} \times \mathcal{F}$ such that
$(r,f)\in \Phi_R$ if room $r$ has feature $f$ and $(e,f)\in \Phi_E$ if
event $e$ requires feature $f$, respectively. Each room $r\in
\mathcal{R}$ has a fixed capacity $C_r$, expressed in terms of seats
for students.

In addition, it is defined a precedence relation $\Pi \subseteq \mathcal{E} \times
\mathcal{E}$, such that if $(e_1,e_2) \in \Pi$, 
events $e_1$ and $e_2$ must be scheduled in timeslots $t_1$ and $t_2$
such that $t_1 < t_2$.

Finally, there is an availability relation $\mathcal{A} \subseteq \mathcal{E} \times
\mathcal{T}$, stating that event $e$ can be scheduled in timeslot $t$ only if
$(e,t) \in \mathcal{A} $.

The (hard) constraints of the problem are the following
ones:

\begin{enumerate}[H1.]
\item \textsf{Conflicts:} Events that share common students cannot be
  scheduled in the same timeslot.
\item \textsf{Compatibility:} An event cannot be allocated in a room
  that is missing one of the features needed by the event, or in a room
  whose capacity is less than the number of students attending the
  event.
\item \textsf{Occupancy:} No more than one event per room per timeslot
  is allowed.
\item \textsf{Availability:} Timeslots must be assigned to events
  according to the availability relation $\mathcal{A}$.
\item \textsf{Precedences:} Timeslots must be assigned to events
  according to the precedence relation $\Pi$.
\end{enumerate}

Since reaching feasibility could be non trivial for some instances, the definition given for ITC
2007 includes the distinction between \emph{valid} and 
\emph{feasible} timetables \cite[see][]{LePM07}.  In a valid timetable all
hard constraints must be satisfied, but it is allowed to leave some events unscheduled (i.e., they have no timeslot assigned). A feasible
timetable is a valid one in which all events are scheduled.

The prescription of the ITC 2007 rules require all solutions to be
valid, but they do allow also infeasible solutions.  In formal terms,
this means that the problem consists in finding an assignment
$\mathcal{E} \rightarrow \mathcal{T} \times \mathcal{R} \cup
\{(t_{\delta}, r_{\delta})\}$, where $t_\delta$ and $r_\delta$ are a
\emph{dummy timeslot} and a \emph{dummy room}. The assignment of an
event to these special entities identifies the unscheduled events. In
addition, we introduce a new hard constraint type, that accounts for
the unscheduled events, which can be violated to some extent:

\begin{enumerate}[H6.]
\item \textsf{Unscheduled Events:} Events cannot be
  unscheduled.\label{con:unscheduled}
\end{enumerate}

The (integer-valued) objective function is the sum
of the following soft constraints. Each violation of any of
the three kinds accounts as one point in the objective function.

\begin{enumerate}[S1.]
\item \textsf{Late Events:} A student should not attend an event in
  the last timeslot of a day. For each event in the last timeslot, we
  compute the sum of the number of students that have to attend
  it.\label{obj:late-event}
\item \textsf{Consecutive Events:} A student should not attend more
  than two consecutive events in a day (i.e., the last timeslot of a
  day and the first one of the following day are not considered as
  consecutive). For each day and for each student, we compute the sum
  of the consecutive events subsequent to the second.  For instance,
  if 3 students have to attend 4 consecutive events in a day, the penalty is
  $3\cdot(4-2) = 6$.  \label{obj:consecutive-event}
\item \textsf{Isolated Events:} A student should not attend only one
  single event in the whole day. For each day, we sum the number of
  students that have to attend isolated
  events.\label{obj:isolated-event}
\end{enumerate}

In conclusion, the quality of the solution is evaluated with an
evaluation function that is composed by two measures: the
\emph{distance to feasibility} (H6) and the \emph{objective function}
(S1 + S2 + S3). The distance to feasibility is computed as the sum of
the numbers of students that require unscheduled events.

The evaluation function is hierarchical, in the sense that valid solutions with the lower distance to feasibility are better solution. If two valid solutions have the same distance
to feasibility, then the solution with the minimum value of the
objective function is preferred.

\subsection{Problem Variants and Available Instances for PE-CTT}
\label{sec:instances}

The problem formulation presented above is the one defined by
\citet{LePM07} and used in the ITC 2007. Two other versions have been
considered in the literature, which are obtained from the above one by
removing some of the constraints.

In particular, the first one is the original formulation, proposed by
the Metaheuristics Network \citep{RSBC03} and used for the ITC 2002.
This formulation does not consider \textsf{Availability} and
\textsf{Precedences}. In addition, since for the ITC 2002 instances
the feasibility was easy to be obtained for all instances, the
possibility to leave some events unscheduled was not taken into account.

\begin{table}
  \centering
  \footnotesize
  \begin{tabular}{|l|*{6}{c}|ccc|} 
    \hline
    Formulation           & H1 & H2 & H3 & H4 & H5 & H6 & S1 & S2 & S3 \\ \hline
    \textsc{Full} (ITC 2007)       &  $\surd$ &  $\surd$ & $\surd$ & $\surd$ & $\surd$ & $\surd$ & $\surd$ & $\surd$ & $\surd$\\
    \textsc{Original} (ITC 2002) &  $\surd$ &  $\surd$ & $\surd$ & --- & --- &  --- & $\surd$ & $\surd$ & $\surd$ \\
    \textsc{Hard-Only}             &  $\surd$ &  $\surd$ & $\surd$ & --- & --- & $\surd$ & --- & --- & --- \\ \hline
  \end{tabular}
  \caption{PE-CTT formulations.}
  \label{tab:formulations}
\end{table}

The other formulation has been proposed by \cite{LePa05}, and is a
further simplification, as it does not include
\textsf{Availability} and \textsf{Precedences} and it discards
all soft constraints. Differently from the previous formulation,
however, it considers the possibility of having unscheduled events.
The formulations considered are summarised in
Table~\ref{tab:formulations}.

Four sets of instances are publicly available and have been used in
the experimental analyses reported in the scientific literature so far.
Table~\ref{tab:instances} lists for each set of instances the origin,
the web site from which they can be downloaded, the formulation
considered, the number of instances that compose the data set, and the
year of publication.

\begin{table}
  \centering
  \footnotesize
  \begin{tabular}{|lp{0.4\textwidth}cc|}\hline
    Instance Family  & Formulation  & \# Instances & Year \\ \hline
    ITC 2007 & \textsc{Full} & 24 & 2007 \\
    \multicolumn{4}{|p{0.5\textwidth}|}{\url{http://www.cs.qub.ac.uk/itc2007}}  \\
    Lewis \& Paechter  & \textsc{Hard-Only} & 60 & 2005\\ 
    \multicolumn{4}{|p{0.9\textwidth}|}{\url{http://www.soc.napier.ac.uk/~benp/centre/timetabling/harderinstances.htm}} \\
    ITC 2002 & \textsc{Original}  & 20 & 2002 \\
    \multicolumn{4}{|p{0.6\textwidth}|}{\url{http://www.idsia.ch/Files/ttcomp2002}} \\
    Metaheuristics Network & \textsc{Original} & 12 & 2001 \\ 
    \multicolumn{4}{|p{0.65\textwidth}|}{\url{http://iridia.ulb.ac.be/supp/IridiaSupp2002-001}} \\
    \hline
  \end{tabular}
  \caption{Available instances for PE-CTT.}
  \label{tab:instances}
\end{table}

All instances are artificial, as they are created by a random
generator, based on realistic bounds for the problem features. For all
of them, the set of timeslots is fixed to $T = 45$, split in 5 days $D = 5$ of $9$ timeslots
each, such that timeslots $\{1,\dots,9\}$ belong to day 1, timeslots
$\{10,\dots,18\}$ belong to day 2, and so on.

Each instance is available in a single text-only file (for the sake of
brevity, we do not report the format here). Two different file formats
are used: one for the \textsc{Full} formulation, which includes
\textsf{Availability} and \textsf{Precedences}, and the other for the
\textsc{Original} and \textsc{Hard-Only} formulations, without them.  In addition, the proposers have released a validator for both
the \textsc{Original} and the \textsc{Full} formulations. We have used it
for certifying the solution quality of all the results we have found
in the experimental phase.

This means that for example the instances of \citet{LePa05} could be
used also for the \textsc{Original} formulation. However, we consider
only the pairs Instance/Formulation that have been investigated in the
past, so that we can compare with previous work.

\section{Related Work}

In the last forty years, starting with \citet{Gotl63}, many papers
related to educational timetabling have been published and several
applications have been developed and employed in practise.  In
addition, many research surveys have been published, going from
\citet{Werr85}, to \citet{Scha99}, to the most recent ones by
\cite{BuPe02} and \cite{Lewi08}. We refer to them for an
introduction to educational timetabling.

With specific regard to course timetabling, the most seminal
works on course timetabling are those by
\citet{Hert91,Hert92}, who uses a Tabu Search approach to solve two
different versions of the problem. More recently, 
\citet{MuMR07} tackle a very complex formulation of the problem and
solve it by decomposition and constraint-based local search. 

However, most of the recent work on course timetabling, besides the
one on PE-CTT, has focused on the other ``standard''
formulation, namely CB-CTT.  To this regard, \citet{LuHa09} solve the
CB-CTT problem by Tabu Search on a large neighbourhood. \citet{LaLu10}
and \citet{BMPR10} both use a IP approach and find several lower
bounds along with a few optimal solutions.  \citet{HaBe11} use a
decomposition approach to improve on the lower bounds obtained with
the model of \citet{LaLu10}.  \cite{Mull09} uses a multi-step local
search approach to find good solutions of the problem. Finally, our
research group \citep{BeDS11} has proposed a hybrid Tabu
Search/Simulated Annealing approach for this problem.

The initial work on PE-CTT has been carried out inside the
Metaheuristics Network by \citet{RSBC03} who compare several
metaheuristic techniques for the \textsc{Original} formulation on
the set of instances specifically designed for their study. That work
has been extended by \citet{CBSR06} that apply the same techniques,
suitably refined and tuned, to the instances defined for the ITC 2002.

The same formulation has been tackled by \citet{Kost04} using a
multi-stage metaheuristic approach. Both \citeauthor{Kost04} and
\citeauthor{CBSR06} consider a search space composed by assignments of
events to timeslots only, leaving the rooms unassigned. The room
assignment is performed by a specialised sub-solver that applies a
matching algorithm. Finally, the ITC 2002 instances have been tested
also by \citet{BBNP03} and \citet{DiSc06}, who also used local search
based techniques.

Moving the \textsc{Full} ITC 2007 formulation, \citet{ChFH08} propose
a solver, built on the previous work for ITC 2002 \citep{CBSR06},
which consists of several heuristic modules in a two phase solution
process (dealing with hard and soft constraints, respectively). The
modules have been assembled and tuned using the automated algorithm
configuration procedure ParamILS \citep{HHLS09}.  \Citet{BrHu10}
design a deterministic heuristic approach that builds a LP solution
using column generation and then tries to improve it by solving ILP
subproblems.  \Citet{Lewi10} employs a three stages strategy in which
a constructive phase is followed by two separate phases of Simulated
Annealing.  The idea behind this method is to arrange constraints
corresponding to different levels of importance in the different
phases of the solution process.

\Citet{Mull09} applies a constraint-based framework incorporating a
series of algorithms based on local search techniques, that operates
over feasible (but not necessarily complete) solutions. Finally,
\citet{CHOP10} proposes both a constraint-based technique and a
multi-stage local search one. This latter method has been the winner
of the PE-CTT track of the ITC 2007.

A few authors considered the \textsc{Hard-Only} formulation and the
corresponding instances. The first works by \citet{LePa05, LePa07} use
an evolutionary algorithms to tackle the problem. Subsequently,
\citet{TuBM07} use a graph-based heuristic to construct a feasible
solution of the relaxed problem (where constraint H2 is partially
relaxed) and then apply a SA-approach relying on a Kempe chains
neighbourhood. Finally, \citet{LiZC11} propose a clique-based
heuristic that tries to identify cliques as the set of events that can
be scheduled in the same timeslot.

\section{Local Search for Post-Enrolment Course Timetabling}

We describe our local search technique in six stages by highlighting
the different components of our solution method. Namely these
components are: preprocessing and constraint reformulation,
search space, initial solution, neighbourhood relations, cost
function, and the Simulated Annealing metaheuristic.

\subsection{Preprocessing and constraint reformulation}

By a careful analysis of the features and the constraints of the
problem it is possible to identify some preliminary preprocessing and
reformulation steps that can significantly improve the efficiency of
the local search phase. This stage is composed by five steps. The
first three steps have already been proposed and employed in previous
works \citep[see, e.g.][]{Kost04}. Instead, the remaining two steps
are our original ideas and they have a substantial impact on the
search strategy of our solver.

\begin{description}

\item[1. Creation of auxiliary matrices:] According to the features held by
  the rooms, the room capacities, and the features requested by the
  events, we create a Boolean-valued \emph{event-room compatibility matrix} $\Theta_R$,
  which states whether a room is suitable for an event. 
  The data about features and capacities can then
  be discarded and replaced completely by the $\Theta_R$ matrix.

  Similarly, according to the student enrolment data we create a symmetric Boolean-valued
  \emph{event-event conflict matrix} $\Theta_E$, which 
  accounts for the presence of common students between pair of events.

\item[2. Propagation of precedences:] Given the precedence relation
  $\Pi$, we perform a preliminary constraint propagation in order to
  restrict the availability for the events.  For any pair of events
  $e_1$ and $e_2$ such that $(e_1,e_2)\in \Pi$, we mark as unavailable
  period $T$ for $e_1$, and period $1$ for $e_2$.  Pursuing further
  this idea, we consider all chains of events (also longer than two)
  in the graph obtained by the transitive closure of the precedence
  relation.  Based on this process (known as arc-consistency in the
  constraint programming community), we determine for each event $e$ a
  minimum and a maximum assignable timeslot. The values outside this
  interval are considered unavailable for $e$, and thus removed from
  the availability relation $\mathcal{A}$.

\item[3. Identification of 1-room events:] Looking at the $\Theta_R$ matrix, it is
  possible to identify events that are compatible only with a single room. We call these events \emph{1-room events}. Obviously, two 1-room events that share the same
  compatible room $r$ cannot be scheduled in the same timeslot. We
  thus update the $\Theta_E$ matrix adding these new conflicts.

\item[4. Identification of all-room events:] A further look at the $\Theta_R$ matrix allows us to
  identify also the events that are compatible with all the rooms. We call these events \emph{all-room events}.

  For this kind of events it is not necessary to assign a room during
  search, and the actual room can be assigned in a simple post-processing
  phase. Therefore, these events are always assigned to the dummy
  room $r_{\delta}$. However, through the search it is still necessary to guarantee that
  the number of events assigned to each timeslot does not exceed the
  number of rooms $R$. 

\item[5. Sorting rooms by the number of compatible events:] In this step, we
  count for each room the number of events that are compatible with it
  (all-room events are not considered in this phase).  This value
  represents a sort of ``attractiveness'' of the room.  We create a list of
  rooms sorted in ascending order of attractiveness.

  This list will be used in the search in order to assign rooms in
  such a way to leave during search the most attractive rooms
  available for further events to be added.
\end{description}

\subsection{Search Space}

As already mentioned, the \emph{solution space} is composed by all the
assignments of timeslots and rooms to events extended with the pair composed by dummy
timeslot and the dummy room: $\mathcal{E} \rightarrow \left( \mathcal{T} \times \mathcal{R} \cup
\{(t_\delta,r_\delta)\}\right)$.

The \emph{search space} employed by our algorithm is the solution space itself, with some
restrictions.  First, only available timeslots and compatible rooms
can be assigned to each event. In addition, assignments are included
in the search space only if no pair of events are assigned to the same
timeslot and room, and the total number of events assigned to a
timeslot is less than or equal to the number of rooms.  Summarising, all
assignments in the search space do not violate the constraints
\textsf{Compatibility}, \textsf{Availability}, and \textsf{Occupancy}.

On the other hand, \textsf{Conflicts}, \textsf{Precedences}, and
\textsf{Unscheduled} can be violated, and thus they are included in
the cost function.

Finally, in the search space all-room events are always assigned to
the dummy room $r_\delta$, and actual rooms will be assigned during
the post-processing phase.

\subsection{Initial Solution}

For the construction of the initial solution, we propose two different
methods.  The first one, denoted by $I_0$, is a greedy procedure
that assigns each event $e$ to a random timeslot $t$, which is available
for $e$ and is not already assigned to $R$ events.

If a room $r$ compatible with $e$ is free in $t$, the pair $(t,r)$ is
assigned to $e$, otherwise the event is assigned to
$(t_\delta,r_\delta)$. Compatible rooms are visited in order of
ascending attractiveness. 

The second method, denoted by $I_1$, is based on the same idea but
it tries to leave unscheduled as few events as possible.  It proceeds
in the same way, but when no room is available in $t$ for
$e$, $I_1$ draws a new random timeslot. However, being a greedy
procedure, it might happen in a given stage that there
is no room compatible with $e$ in any timeslot. In order to avoid an infinite loop in such a situation, we stop the procedure after a finite number of draws and
assigns $e$ to $(t_\delta,r_\delta)$. 
For example, for the ITC 2007 instances, the number of unscheduled
events of a state generated with $I_1$ is most of the times 0, and
occasionally it is 1 or 2. On the contrary, for $I_0$ up
to 25\% of the events might be left unscheduled in the most difficult
instances.

\subsection{Neighbourhood relations}

Two different neighbourhood relations are considered in this work:

\begin{description}
\item[\textsf{MoveEvent (ME)}:] Move one event $e\in \mathcal{E}$ from its
  currently assigned timeslot to timeslot $t\in \mathcal{T}\cup \{t_\delta\}$.
  The move $\mathsf{ME}(e,t)$ is admissible if $t$ is available for
  $e$ and there is a compatible free room $r$ for $e$ in $t$.
The pair $(t,r)$ is assigned to $e$.

\item[\textsf{SwapEvents (SE)}:] Swap the timeslots $t_1,t_2\in
  \mathcal{T}\cup \{t_\delta\}$ assigned to two events $e_1,e_2 \in
  \mathcal{E}$.  The move $\mathsf{SE}(e_1,e_2)$ is admissible if $t_1
  \neq t_2$ and $t_1$ (resp.\ $t_2$) is available for $e_2$ (resp.\
  $e_1$) and there is a compatible free room $r_1$ (resp.\ $r_2$) for
  $e_2$ (resp.\ $e_1$) in $t_1$ (resp.\ $t_2$). The pair $(t_2,r_2)$
  is assigned to $e_1$ and the pair $(t_1,r_1)$ is assigned to $e_2$.
\end{description}

For both neighbourhoods, rooms are explored in ascending order of
attractiveness. For events in timeslot $t_\delta$ and for all-room events the only room considered compatible is
$r_\delta$.

For the neighbourhood \textsf{ME} we also consider a restricted version
that we call $\mathsf{ME}^-$. The move $\mathsf{ME^-}(e,t)$ is
admissible only if $t \neq t_\delta$. Intuitively, $\mathsf{ME}^-$
excludes the moves that increase the number of unscheduled events in
the current state.

\subsection{Cost Function}
\label{sec:cost-function-pectt}

The cost function that guides the search is a combination of the
soft constraint penalty and the violation of hard constraints. In
detail, \textsf{Compatibility}, \textsf{Occupancy}, and
\textsf{Availability} are always satisfied in the search space,
whereas \textsf{Conflicts}, \textsf{Precedences}, and
\textsf{Unscheduled Events} can be violated, and therefore they are
included in the cost function.

In case of violation of the \textsf{Unscheduled Events} constraint,
the formulation prescribes to count the number of students that are
enrolled in the event. Consequently, in order to have comparable
values also for the other hard constraints components, in case of a
violation of the \textsf{Conflicts} and \textsf{Precedences}
constraints we count the minimum between the number of students of the
two events involved.  However, for the purpose of having at the end of
the run only possible violations of the \textsf{Unscheduled Events}
component (as required), in the last few iterations of the search the
cost of \textsf{Conflicts} and \textsf{Precedences} is doubled. This
proved experimentally to be sufficient to ensure that there are no
violations of a type different from \textsf{Unscheduled Events}.

In conclusion, the cost function $F$ is the composition of two
terms: the distance to feasibility, multiplied by a suitable high
weight $W$, and the objective function $f$.

Given that we make one single step and that the move acceptance is
based on $\Delta F$, the value of $W$ is crucial of the performances
of our solver. In fact, if $W$ is too high the start temperature needs
to be set to a very high value, which would result if a waste of time
for the search.  On the other hand, if $W$ is too small it is possible
that the solver follows trajectories that ``prefer'' infeasible
solutions to feasible ones, if they have lower objective cost. In
conclusion, $W$ needs to be set experimentally, as discussed in
Section~\ref{sec:exper-analys}.

\subsection{Simulated Annealing}

Many versions of SA have been proposed in the literature \citep[see,
e.g.,][]{KiGV83,Egle90,AaLe97,HoSt05}. The version used here is
the one with probabilistic acceptance and geometric cooling. In detail, at each
iteration of the search process a random neighbour is selected. The
move is performed either if it is an improving one or according to an
exponential time-decreasing probability. If the cost of the
move is $\Delta F > 0$, the move is accepted with probability
$e^{-\Delta F/ T}$, where $T$ is a time-decreasing parameter called
\emph{temperature}. At each temperature level a number $N$ of
neighbours of the current solution is sampled and the new solution is
accepted according the above mentioned probability distribution.  The
value of $T$ is modified using a \emph{geometric} schedule, i.e.,
$T_{i+1} = \beta \cdot T_i$, in which the parameter $\beta < 1$ is called
the \emph{cooling rate}.  The search starts at temperature $T_0$ and
stops when it reaches $T_{min}$.

Different settings of the parameters of SA would result in different
running times. Instead, we want to compare them in a fair
setting, giving to all of them the same amount of computational time.
To this aim, we let the three parameters $T_0$, $T_{min}$, and $\beta$
vary and we fix $N$ in such a way to have exactly the same number of
total iterations. Calling $I$ the fixed total number of
iterations, we compute $N$ from the following formula.

$$N = I / \log_{\beta}{(T_0 / T_{min})} $$

In this way, the total running time is approximately the same, for all
combinations of parameters.

We experiment with three solvers, that differ from each other based on
the neighbourhood used and the initial solution procedure. The first
solver we consider is SA using as neighbourhood the union of
$\mathsf{ME}$ and $\mathsf{SE}$, and  as the initial state method $I_0$.
We denote it as $SA(I_{0}, \mathsf{ME} \oplus \mathsf{SE})$. Using a similar notation, the
other two solvers are denoted by $SA(I_{0}, \mathsf{ME^-}\oplus\mathsf{SE})$ and
$SA(I_{1}, \mathsf{ME^-}\oplus \mathsf{SE})$.

Intuitively, the first one explores freely the full search space,
composed also by unscheduled events. The second one starts with a
state with unscheduled events, but leads the search as much as
possible in the direction of feasible solutions. The third one
focuses on the space in which all events are scheduled.

The total number of iterations is set to $I = 1.14\cdot 10^8$, which
corresponds approximately to the time granted for ITC 2007 and which
results in a running time of about 300s on our PC, an Intel Core i7 @1.6 GHz
PC (64 bit). We prefer to set the number of iterations, rather than
using a real timeout because, as advocated by \citet{John96}, the use
of the timeout makes the experiments less reproducible.

The software is written in C++ language, it uses the framework
\textsc{EasyLocal++} \citep{DiSc03b}, and it is compiled using the GNU
C/C++ compiler, v. 4.4.3, under Ubuntu Linux. 

\section{Experimental Analysis}
\label{sec:exper-analys}

For tuning the three solvers, we first select the parameters to be
evaluated.  To this regard, we decide to use $T_0$, $\beta$, and $\rho
= T_0/T_{min}$, which turned out to provide a better selection of the
configurations than using $T_{min}$ directly. Given that we use two
different types of moves, namely $\mathsf{ME}$ (or $\mathsf{ME^-}$)
and $\mathsf{SE}$, we add an additional parameter, called $sr$ (for
swap rate) which is the probability of drawing a move of type
$\mathsf{SE}$. Finally, the parameter $W$ needs to be set.

\subsection{Preliminary Screening}

In order to perform an effective tuning it is useful to have a screening
based on preliminary experiments, that allows us to eliminate some of
the five parameters and to focus on the most important ones.

Preliminary experiments show that $\beta$ is not significant. This is
not surprising, because in our setting $N$ is a function of the other
parameters, and therefore $\beta$ only determines the entity of the
single step in the temperature and not the actual slope of the cooling
trajectory, which is determined by $\rho$. We therefore set $\beta$
to the fixed value $0.9999$.

Preliminary experiments show also that $sr$ is not significant, as
long as it is set within the range $[0.1,0.5]$. Consequently, we fixed
$sr$ to the value $0.4$, which provided marginally better results.

Regarding $W$, it turned out that the value $W = 1$ is big enough 
to ensure that the solver prefers feasible solutions to infeasible ones.
Therefore this parameter is set to 1 for all experiments.
This surprising finding is explained by the observation that a hard violations 
has a cost equal the number of students involved, whereas only a fraction of the 
students involved in the move contributes to the soft constraints.

\subsection{Experimental Design and Tuning}

For the remaining two parameters ($T_0$ and $\rho$), we have to select
the configurations to be tested. Instead of using a classical
\emph{full factorial} design, which consists in a regular sampling of the range for
each parameter and testing all combinations, we resort to the
\emph{Nearly Orthogonal Latin Hypercubes (NOLH)} proposed by
\citet{CiLu07}, that allow us to fill the space using much less
configurations. To generate the actual configurations we use the NOLH
spreadsheet made available by \citet{Sanc05}, using the design with 33
points, within the ranges $T_0 \in [1,100]$ and $\rho \in [10,1000]$.

For the comparison of the 33 configurations we resort to $F$-Race
\citep{Bira05}, which is a sequential testing procedure
that uses the Friedman two-way analysis of variance by ranks to decide
upon the elimination of inferior candidates. At each stage, a new
instance is selected, all remaining configurations run on it, and weaker
configurations are discarded if enough statistical evidence has arisen
against them.  We use the canonical value $0.05$ as significance level in the
tests.  The transformation of results in ranks prescribed in $F$-Race
guarantees that in the statistical procedure the aggregation of
results over the instances is not influenced by the
differences in the scale of the cost function values that depends on the instance at hand.

Considering that each set of instances has different features and they
refer to different problem formulations (see
Table~\ref{tab:instances}), we decide to tune separately the
parameters for each instance family.
A tuning process directed to a general and unique parameter setting 
is also possible and it leads to only slightly inferior results, 
proving that the algorithm is robust enough. 

For each instance family and for each solver we firstly
compare the 33 configurations resulting from the NOLH analysis,
obtaining the best parameter configuration for each solver. Then, for
each instance family, we compare the three solvers using their best
configuration by means of the \emph{Wilcoxon rank-sum test}. 

Table~\ref{tab:comp_solvers} reports, for each family, the best
solver, along with its best configuration. There are cases in which
the difference between solvers or configurations is not statistically
significant. In these situations, we consider as the best the one with
minimum average rank.

In order to get close to the setting of the ITC 2007, for the
\textsc{Full} formulation we use the instances 1-16 for tuning the
parameters, and the instances 17-24 for validation. In fact, for the
competition the last 8 instances (Hidden Instances) where not given to
the participants, but used by the organisers for evaluating the
solvers submitted.

\begin{table}
\centering
\footnotesize
\begin{tabular}{|r@{}l|l|r|r|}\hline
\multicolumn{2}{|c|}{Instance Family} & \multicolumn{1}{c|}{Solver} & \multicolumn{1}{c|}{$T_0$} & \multicolumn{1}{c|}{$\rho$} \\ \hline
\multicolumn{2}{|l|}{ITC 2007} & $SA(I_{0}, \mathsf{ME^-}\oplus\mathsf{SE})$ & 20.41  & 33.88 \\
 \multirow{2}{*}{Lewis \& Paechter}  &~~ Med  & $SA(I_{0}, \mathsf{ME}\oplus\mathsf{SE})$ & 31.62 & 257.63 \\
 						     &~~ Big & $SA(I_{0}, \mathsf{ME}\oplus\mathsf{SE})$ & 36.30 & 295.12 \\
\multicolumn{2}{|l|}{ITC 2002} & $SA(I_{1}, \mathsf{ME^-}\oplus \mathsf{SE})$	 & 3.89 &  31.62  \\
\multicolumn{2}{|l|}{Metaheuristics Network} & $SA(I_{0}, \mathsf{ME}\oplus\mathsf{SE})$  & 3.89 & 31.62  \\
\hline
\end{tabular}
\caption{Best settings of the SA equipped solvers.}
\label{tab:comp_solvers}
\end{table}
\subsection{Comparison with Best Known Results}

We now compare the solvers emerged from the tuning phase
(Table~\ref{tab:comp_solvers}) with the best results in the
literature. Table~\ref{tab:solvers} summarises the solvers with which
we compare. For each solver, we report the reference, the techniques
used and the family of instances it solves. Notice that no solver previously presented in the literature has dealt with more than one family of instances.

\subsubsection{ITC 2007 instances}

\begin{sidewaystable}
\footnotesize
\begin{tabular}{|c|l|l|l|}\hline
Solver & Reference & Technique & Family of instances \\ \hline
A & \citet{ChFH08} & Local Search + Matching  & ITC 2007 \\
B & \citet{Mull09} &Constructive + Local Search (HC, GD, SA) & ITC 2007\\
C1 & \citet{CHOP10}& Local Search (SA) & ITC 2007 \\
C2 & \citet{CHOP10}& Local Search (SA) & ITC 2007 \\
D & \citet{Lewi10} & Constructive + Iterated Heuristic + Local Search (SA) & ITC 2007 \\
E & \citet{BrHu10} & Column Generation + ILP & ITC 2007 \\
F & \citet{MNCR08}& Ant Colony Optimisation & ITC 2007, ITC 2002 \\
G1 & \citet{LePa07}& Genetic Algorithm & \citeauthor{LePa07} \\
G2 & \citet{LePa07}& Genetic Algorithm & \citeauthor{LePa07} \\
G3 & \citet{LePa07} & Genetic Algorithm & \citeauthor{LePa07} \\
H & \citet{TuBM07} & Constructive + Local Search (SA) & \citeauthor{LePa07} \\
I & \citet{LiZC11} &  Constructive & \citeauthor{LePa07} \\
J & \citet{BBNP03} & Local Search (GD) & ITC 2002 \\
K & \citet{DiSc06} & Local Search & ITC 2002 \\
L & \citet{Kost04} & Constructive + Local Search (SA) & ITC 2002 \\
M & \citet{CBSR06} & Constructive + Local Search (TS, SA) & ITC 2002 \\
N & \citet{SoKS02} & Ant Colony Optimisation & Metaheuristics Network \\
O & \citet{AbBM07b} & Memetic algorithm & Metaheuristics Network\\
P & \citet{McMu07} & Constructive + Local Search (GD) & Metaheuristics Network \\
Q & \citet{LaOb08} & Local Search (GD) & Metaheuristics Network \\
R & \citet{TuSM09} & Local Search (GD) & Metaheuristics Network \\ 
\hline
\end{tabular}
\caption{Solvers compared in the experimental phase (HC: Hill Climbing, GD: Great Deluge, SA: Simulated Annealing, TS: Tabu Search.)}
\label{tab:solvers}
\end{sidewaystable}

We first consider the ITC 2007 instances.
Table~\ref{tab:ITC2007_results} reports the values obtained by our
method in 30 runs for instances 1--16, along with a comparison with
respect to the available results reported in the literature (in bold
the best results). The presence of the  dash symbol means that no feasible solution has been found.

The columns \%Feas show the percentage of feasible
solutions obtained. Notice that, for solver D, the paper in some cases
reports only that this percentage is greater than 95\%, instead of the
precise value (it reports instead the average number of violations).

The average values are computed considering all the solutions obtained
in the experiments, including the infeasible ones. Obviously, the
value of the objective function for the infeasible solutions is not
very meaningful. However, for our solver the number of infeasible
solutions is very small, therefore the average of the value of the
objective function is still the most meaningful index.

For instances 17--24, values are not reported in the cited papers
\citep[except for][]{BrHu10}, therefore we compare our solver with the results
extracted from the spreadsheet available from the ITC 2007 website. 
As mentioned above, our results on these instances are obtained with the best parameter configuration used for instances 1--16. 

From the results it is possible to see that our method outperforms all
other solvers on 9 out of 24 instances, it is second to
\citeauthor{CHOP10} on 11 instances, and second to \citeauthor{MNCR08}
in the remaining 4.

This positive result is confirmed by applying the ranking method of
ITC 2007. The first row of Table~\ref{tab:rankITC2007} shows the
average of the ranks obtained by each finalist of the competition
(available from the ITC 2007 website), from which it results that
Cambazard \emph{et al.} won the competition. Adding a-posteriori our
solver to the final of the competition\footnote{Using the spreadsheet
  downloaded from the ITC 2007 website.}, we obtain the ranks of the
second row, from which we see that our solver would have won the
competition.

\begin{sidewaystable}
\centering
\footnotesize
\subfloat[Public instances\label{tab:ITC2007_results_public}]{
\begin{tabular}{|c|r|r|rrr|rrr|rrr|r|rrr|rrr|}
\hline
& \multicolumn{1}{c|}{A} & \multicolumn{1}{c|}{B} & \multicolumn{3}{c|}{C1} & \multicolumn{3}{c|}{C2} & \multicolumn{3}{c|}{D} & \multicolumn{1}{c|}{E} & \multicolumn{3}{c|}{F} & \multicolumn{3}{c|}{Us}  \\ 
Inst. & \multicolumn{1}{c|}{Best} & \multicolumn{1}{c|}{Best} & \multicolumn{1}{c}{Avg} & \multicolumn{1}{c}{\%Feas} & \multicolumn{1}{c|}{Best} &  \multicolumn{1}{c}{Avg} & \multicolumn{1}{c}{\%Feas} & \multicolumn{1}{c|}{Best} &  \multicolumn{1}{c}{Med} & \multicolumn{1}{c}{\%Feas} & \multicolumn{1}{c|}{Best} &  \multicolumn{1}{c|}{Best} & \multicolumn{1}{c}{Avg} & \multicolumn{1}{c}{\%Feas} & \multicolumn{1}{c|}{Best} & \multicolumn{1}{c}{Avg} & \multicolumn{1}{c}{\%Feas} & \multicolumn{1}{c|}{Best}  \\ \hline
1 & 925 & 1330 & 830 & 100 & 358 & 547 & 100 & 15 & 1492 & 45 & 1294 & 1636 & 613 & 54 & 0 & \textbf{399.2} & 100 & 59 \\ 
2 & 1156 & 2154 & 924 & 100 & 11 & 403 & 100 & 356 & 1826 & 22 & 1599 & 1634 & 556 & 59 & 0 & \textbf{142.2} & 100 & 0 \\ 
3 & 179 & 205 & 224 & 100 & 156 & 254 & 100 & 174 & 457 & $>95$ & 278 & 355 & 680 & 100 & 110 & \textbf{209.9} & 100 & 148 \\ 
4 & 66 & 394 & 352 & 100 & 61 & 361 & 100 & 249 & 589 & $>95$ & 388 & 644 & 580 & 100 & 53 & \textbf{349.6} & 100 & 25 \\ 
5 & 52 & 0 & \textbf{3} & 100 & 0 & 26 & 100 & 0 & 193 & $>95$ & 22 & 525 & 92 & 100 & 13 & 7.7 & 100 & 0 \\ 
6 & 536 & 13 & 14 & 100 & 0 & 16 & 100 & 0 & 689 & $>95$ & 369 & 640 & 212 & 95 & 0 & \textbf{8.6} & 100 & 0 \\ 
7 & 7 & 5 & 11 & 100 & 5 & 8 & 100 & 1 & 421 & $>95$ & 74 & 0 & \textbf{4} & 100 & 0 & 4.9 & 100 & 0 \\ 
8 & 0 & 0 & \textbf{0} & 100 & 0 & \textbf{0} & 100 & 0 & 206 & 100 & 0 & 241 & 61 & 100 & 0 & 1.5 & 100 & 0 \\ 
9 & 1480 & 1895 & 1649 & 100 & 1049 & 1167 & 100 & 29 & 2312 & 2 & 1482 & 1889 & \textbf{202} & 85 & 0 & 258.8 & 97 & 0 \\ 
10 & 1364 & - & 2003 & 98 & 773 & 1297 & 89 & 2 & 2262 & 2 & 2380 & 1677 & \textbf{4} & 100 & 0 & 186.4 & 97 & 3 \\ 
11 & 166 & 347 & 311 & 100 & 157 & 361 & 100 & 178 & 541 & $>95$ & 344 & 615 & 774 & 99 & 143 & \textbf{269.5} & 100 & 142 \\ 
12 & 1 & 453 & 408 & 100 & 0 & \textbf{380} & 100 & 14 & 741 & $>95$ & 486 & 528 & 538 & 86 & 0 & 400.0 & 100 & 267 \\ 
13 & 360 & 74 & \textbf{89} & 100 & 0 & 135 & 100 & 0 & 631 & $>95$ & 365 & 485 & 360 & 94 & 5 & 120.0 & 100 & 1 \\ 
14 & 576 & 2 & \textbf{1} & 100 & 0 & 15 & 100 & 0 & 660 & $>95$ & 222 & 739 & 41 & 100 & 0 & 3.6 & 100 & 0 \\ 
15 & 0 & 0 & 80 & 100 & 0 & 47 & 100 & 0 & 344 & $>95$ & 266 & 330 & \textbf{29} & 100 & 0 & 48.0 & 100 & 0 \\ 
16 & 0 & 6 & \textbf{19} & 100 & 1 & 58 & 100 & 1 & 194 & $>95$ & 99 & 260 & 101 & 100 & 0 & 50.1 & 100 & 0 \\ \hline
Avg &  &  & 432.4 &  &  & 317.2 &  &  & 847.4 &  &  &  & 302.9 &  &  & \textbf{153.7} &  &  \\ \hline
\end{tabular}
}

\subfloat[Hidden instances\label{tab:ITC2007_results_hidden} (results taken from the spreadsheet on the ITC 2007 website)]{
\begin{tabular}{|c|rr|rrr|rrr|r|rrr|rrr|}
\hline
\multicolumn{1}{|c|}{} & \multicolumn{ 2}{c|}{A} & \multicolumn{ 3}{c|}{B} & \multicolumn{ 3}{c|}{C1} & \multicolumn{1}{c|}{E} & \multicolumn{3}{c|}{F} & \multicolumn{ 3}{c|}{Us} \\ 
\multicolumn{1}{|c|}{Inst.} & \multicolumn{1}{c}{Avg} & \multicolumn{1}{c|}{Best} & \multicolumn{1}{c}{Avg} & \multicolumn{1}{c}{\%Feas} & \multicolumn{1}{c|}{Best} & \multicolumn{1}{c}{Avg} & \multicolumn{1}{c}{\%Feas} & \multicolumn{1}{c|}{Best} & \multicolumn{1}{c|}{Best} & \multicolumn{1}{c}{Avg} & \multicolumn{1}{c}{\%Feas} & \multicolumn{1}{c|}{Best}  & \multicolumn{1}{c}{Avg} & \multicolumn{1}{c}{\%Feas} & \multicolumn{1}{c|}{Best} \\ \hline
17 & 9.8 & 5 & 106.2 & 100 & 72 & 4.9 & 100 & 0 & 35 & 116.4 & 100 & 68 & \textbf{0.0} & 100 & 0 \\ 
18 & 339.9 & 3 & 314.3 & 100 & 70 & \textbf{14.1} & 100 & 0 & 503 & 264.8 & 100 & 26 & 41.1 & 100 & 0 \\ 
19 & 2080.8 & 1869 & 2314.0 & 0 &  --  & 2027.0 & 20 & 1824 & 963 & \textbf{233.1} & 90 & 22 & 951.5 & 74 & 0 \\ 
20 & 640.5 & 596 & 919.3 & 100 & 878 & \textbf{505.0} & 100 & 445 & 1229 & -- & 0 & -- & 700.2 & 100 & 543 \\ 
21 & 876.3 & 602 & 336.8 & 100 & 40 & \textbf{27.1} & 100 & 0 & 670 & 326.6 & 80 & 33 & 35.9 & 97 & 5 \\ 
22 & 1839.2 & 1364 & 1593.7 & 60 & 889 & 550.8 & 90 & 29 & 1956 & 82.7 & 100 & 0 & \textbf{19.9} & 100 & 5 \\ 
23 & 1043.4 & 688 & 701.3 & 100 & 436 & \textbf{330.5}& 100 & 238 & 2368 & -- & 0 & -- & 1707.7 & 20 & 1292 \\ 
24 & 963.4 & 822 & 518.0 & 100 & 372 & 124.2 & 100 & 21 & 945 & 129.2 & 100 & 30 & \textbf{105.3} & 100 & 0 \\  \hline
Avg & 974.2 &  & 850.5 &  &  & 448.0 &  &  &  & -- &  &  & \textbf{445.2} &  &  \\ \hline
\end{tabular}
}
\caption{Results on ITC 2007 instances for 30 runs.}
\label{tab:ITC2007_results}
\end{sidewaystable}

\begin{table}[htdp]
\begin{center}
\begin{footnotesize}
\begin{tabular}{|c|c|c|c|c|c|}
\hline
Atsuta \emph{et al.} & C1 & A & F & B & Us \\ \hline
24.43 & \textbf{13.90} & 28.34 & 29.52 & 31.31 & \\ \hline
31.41 & 19.98 & 36.85 & 37.33 & 40.70 & \textbf{16.73} \\ \hline
\end{tabular}
\end{footnotesize}
\end{center}
\caption{Comparison using the ITC 2007 ranking system.}
\label{tab:rankITC2007}
\end{table}

\subsubsection{\citeauthor{LePa05} instances}

Moving to the \citeauthor{LePa05} instances, Tables
\ref{tab:Lewis_media_results} and~\ref{tab:Lewis_big_results} report
the results for the 20 medium and the 20 big instances (we do not
report here results on the 20 small ones because they are not
challenging, given that we solve all of them to optimality in all
runs). 

Following \citet{LePa07}, for these instances in
Tables~\ref{tab:Lewis_results} we report the number of unscheduled
events, rather than the total number of students attending them.
However, the solver, similarly to \citet{TuBM07}, still uses the
number of students as the distance to feasibility. This version of the
cost function proved experimentally to be more effective.

\begin{table}
\centering
\scriptsize

\subfloat[Medium instances\label{tab:Lewis_media_results}]{
\begin{tabular}{|c|r|r|r|rr|rr|rr|}\hline
\multicolumn{1}{|l|}{} & \multicolumn{1}{c|}{G1} & \multicolumn{1}{c|}{G2} & \multicolumn{1}{c|}{G3} & \multicolumn{ 2}{c|}{H} & \multicolumn{ 2}{c|}{I} & \multicolumn{ 2}{c|}{Us} \\ 
Inst. & \multicolumn{1}{c|}{Best} & \multicolumn{1}{c|}{Best} & \multicolumn{1}{c|}{Best} & \multicolumn{1}{c}{Avg} & \multicolumn{1}{c|}{Best} & \multicolumn{1}{c}{Avg} & \multicolumn{1}{c|}{Best} & \multicolumn{1}{c}{Avg} & \multicolumn{1}{c|}{Best} \\ 
\hline
 M1  & 0 & 0 & 0 & \textbf{0.00} & 0 & \textbf{0.00} & 0 & \textbf{0.00} & 0 \\ 
    M2  & 0 & 0 & 0 & \textbf{0.00} & 0 & \textbf{0.00} & 0 & \textbf{0.00} & 0 \\ 
    M3  & 0 & 0 & 0 & \textbf{0.00} & 0 & \textbf{0.00} & 0 & \textbf{0.00} & 0 \\ 
    M4  & 0 & 0 & 0 & \textbf{0.00} & 0 & \textbf{0.00} & 0 & \textbf{0.00} & 0 \\ 
    M5  & 8 & 0 & 0 & \textbf{0.00} & 0 & 0.00 & 0 & \textbf{0.00} & 0 \\ 
    M6  & 15 & 0 & 0 & \textbf{0.00} & 0 & \textbf{0.00} & 0 & 0.90 & 0 \\ 
    M7  & 41 & 34 & 14 & 4.15 & 1 & 3.55 & 0 & \textbf{0.00} & 0 \\ 
    M8  & 21 & 9 & 0 & \textbf{0.00} & 0 & \textbf{0.00} & 0 & 0.30 & 0 \\ 
    M9  & 30 & 17 & 2 & 4.90 & 0 & 2.15 & 0 & \textbf{0.35} & 0 \\ 
    M10  & 0 & 0 & 0 & \textbf{0.00} & 0 & \textbf{0.00} & 0 & \textbf{0.00} & 0 \\ 
    M11  & 12 & 0 & 0 & \textbf{0.00} & 0 & \textbf{0.00} & 0 & \textbf{0.00} & 0 \\ 
    M12  & 0 & 0 & 0 & \textbf{0.00} & 0 & \textbf{0.00} & 0 & 0.60 & 0 \\ 
    M13  & 23 & 3 & 0 & 0.50 & 0 & \textbf{0.00} & 0 & \textbf{0.00} & 0 \\ 
    M14  & 0 & 0 & 0 & \textbf{0.00} & 0 & \textbf{0.00} & 0 & 0.05 & 0 \\ 
    M15  & 10 & 0 & 0 & 0.05 & 0 & \textbf{0.00} & 0 & \textbf{0.00} & 0 \\ 
    M16  & 50 & 30 & 1 & 5.15 & 1 & 0.30 & 0 & \textbf{0.00} & 0 \\ 
    M17  & 21 & 0 & 0 & \textbf{0.00} & 0 & \textbf{0.00} & 0 & 0.15 & 0 \\ 
    M18  & 15 & 0 & 0 & 6.05 & 0 & \textbf{0.00} & 0 & 0.30 & 0 \\ 
    M19  & 51 & 0 & 0 & 5.45 & 0 & \textbf{0.00} & 0 & 0.50 & 0 \\ 
    M20  & 15 & 0 & 3 & 10.60 & 2 & 0.65 & 0 & \textbf{0.55} & 0 \\    \hline
    Avg & & & & 1.84 & & 0.33 & & \textbf{0.19} & \\
    \hline 
\end{tabular}
}

\subfloat[Big instances\label{tab:Lewis_big_results}]{
\begin{tabular}{|c|r|r|r|rr|rr|rr|}\hline
\multicolumn{1}{|l|}{} & \multicolumn{1}{c|}{G1} & \multicolumn{1}{c|}{G2} & \multicolumn{1}{c|}{G3} & \multicolumn{ 2}{c|}{H} & \multicolumn{ 2}{c|}{I} & \multicolumn{ 2}{c|}{Us} \\ 
Inst. & \multicolumn{1}{c|}{Best} & \multicolumn{1}{c|}{Best} & \multicolumn{1}{c|}{Best} & \multicolumn{1}{c}{Avg} & \multicolumn{1}{c|}{Best} & \multicolumn{1}{c}{Avg} & \multicolumn{1}{c|}{Best} & \multicolumn{1}{c}{Avg} & \multicolumn{1}{c|}{Best} \\ \hline
B1  & 0 & 0 & 0 & \textbf{0.00} & 0 & \textbf{0.00} & 0 & 0.15 & 0 \\ 
    B2  & 0 & 0 & 0 & \textbf{0.00} & 0 & \textbf{0.00} & 0 & 0.60 & 0 \\ 
    B3  & 0 & 0 & 0 & \textbf{0.00} & 0 & \textbf{0.00} & 0 & 1.45 & 0 \\ 
    B4  & 32 & 30 & 8 & \textbf{0.00} & 0 & \textbf{0.00} & 0 & \textbf{0.00} & 0 \\ 
    B5  & 31 & 24 & 30 & 1.10 & 0 & 3.20 & 1 & \textbf{0.00} & 0 \\ 
    B6  & 90 & 71 & 77 & 8.45 & 5 & 15.40 & 10 & \textbf{2.85} & 1 \\ 
    B7  & 150 & 145 & 150 & 58.30 & 47 & 46.65 & 39 & \textbf{29.25} & 21 \\ 
    B8  & 35 & 30 & 5 & \textbf{0.00} & 0 & \textbf{0.00} & 0 & \textbf{0.00} & 0 \\ 
    B9  & 26 & 18 & 3 & 0.05 & 0 & \textbf{0.00} & 0 & \textbf{0.00} & 0 \\ 
    B10  & 36 & 32 & 24 & 1.25 & 0 & 1.95 & 0 & \textbf{0.00} & 0 \\ 
    B11  & 43 & 37 & 22 & 0.35 & 0 & 2.35 & 0 & \textbf{0.00} & 0 \\ 
    B12  & 4 & 0 & 0 & \textbf{0.00} & 0 & \textbf{0.00} & 0 & 1.15 & 0 \\ 
    B13  & 23 & 10 & 0 & \textbf{0.00} & 0 & \textbf{0.00} & 0 & 1.15 & 0 \\ 
    B14  & 8 & 0 & 0 & \textbf{0.00} & 0 & \textbf{0.00} & 0 & 1.20 & 0 \\ 
    B15  & 120 & 98 & 0 & \textbf{0.00} & 0 & \textbf{0.00} & 0 & 3.50 & 1 \\ 
    B16  & 120 & 100 & 19 & 2.00 & 0 & \textbf{0.00} & 0 & 0.65 & 0 \\ 
    B17  & 260 & 243 & 163 & 89.90 & 76 & \textbf{2.05} & 0 & 22.00 & 12 \\ 
    B18  & 199 & 173 & 164 & 62.60 & 53 & \textbf{1.70} & 0 & 13.55 & 8 \\ 
    B19  & 262 & 253 & 232 & 127.00 & 109 & 53.20 & 40 & \textbf{52.85} & 37 \\ 
    B20  & 186 & 165 & 149 & 46.70 & 40 & \textbf{14.05} & 9 & 15.05 & 11 \\ 
    \hline
    Avg &&&&19.89 && \textbf{7.03}&&7.27& \\
 \hline
\end{tabular}
}

\caption{Results on \citeauthor{LePa07} instances of for 20 runs.}
\label{tab:Lewis_results}
  
\end{table}

Also for these instances we have been able to obtain new best results
and to be relatively close to the best known results in all the other
cases. It is worth mentioning that these instances have very different
structure with respect to the other data sets, and in these cases it
is not always possible to find a feasible solution. Indeed, the
authors who considered these instances used \emph{ad hoc} techniques,
which are rather different from those used by the authors who worked
on the other instance families.

\subsubsection{ITC 2002 instances}

\begin{table}[t]
 \centering
 \footnotesize
 \begin{tabular}{|c|r|r|r|r|rr|rr|}\hline
 & \multicolumn{1}{c|}{J} & \multicolumn{1}{c|}{K} & \multicolumn{1}{c|}{L} & \multicolumn{1}{c|}{M} & \multicolumn{ 2}{c|}{F} & \multicolumn{ 2}{c|}{Us} \\ 
Inst. & \multicolumn{1}{c|}{Best} & \multicolumn{1}{c|}{Best} & \multicolumn{1}{c|}{Best} & \multicolumn{1}{c|}{Best} & \multicolumn{1}{c}{Avg} & \multicolumn{1}{c|}{Best} & \multicolumn{1}{c}{Avg} & \multicolumn{1}{c|}{Best} \\ 
\hline
1 & 85 & 63 & \textbf{16} & 45 & 82 & 55 & 57.05 & 45 \\ 
2 & 42 & 46 & \textbf{2} & 14 & 64 & 43 & 33.20 & 20 \\ 
3 & 84 & 96 & \textbf{17} & 45 & 92 & 61 & 53.20 & 43 \\ 
4 & 119 & 166 & \textbf{34} & 71 & 208 & 134 & 109.90 & 87 \\ 
5 & 77 & 203 & \textbf{42} & 59 & 185 & 134 & 91.70 & 71 \\ 
6 & 6 & 92 & \textbf{0} & 1 & 59 & 32 & 14.05 & 2 \\ 
7 & 12 & 118 & \textbf{2} & 3 & 138 & 52 & 13.70 & \textbf{2} \\ 
8 & 32 & 66 & \textbf{0} & 1 & 107 & 48 & 20.00 & 9 \\ 
9 & 184 & 51 & \textbf{1} & 8 & 70 & 39 & 21.90 & 15 \\ 
10 & 90 & 81 & \textbf{21} & 52 & 118 & 77 & 60.70 & 41 \\ 
11 & 73 & 65 & \textbf{5} & 30 & 75 & 39 & 38.20 & 24 \\ 
12 & 79 & 119 & \textbf{55} & 75 & 143 & 102 & 83.65 & 62 \\ 
13 & 91 & 160 & \textbf{31} & 55 & 156 & 94 & 77.95 & 59 \\ 
14 & 36 & 197 & \textbf{11} & 18 & 175 & 109 & 34.20 & 21 \\ 
15 & 27 & 114 & \textbf{2} & 8 & 89 & 47 & 11.80 & 6 \\ 
16 & 300 & 38 & \textbf{0} & 55 & 45 & 26 & 16.70 & 6 \\ 
17 & 79 & 212 & \textbf{37} & 46 & 143 & 78 & 56.45 & 42 \\ 
18 & 39 & 40 & \textbf{4} & 24 & 59 & 35 & 25.85 & 11 \\ 
19 & 86 & 185 & \textbf{7} & 33 & 187 & 119 & 72.95 & 56 \\ 
20 & \textbf{0} & 17 & \textbf{0} & \textbf{0} & 38 & 19 & 1.75 & \textbf{0} \\ \hline
Avg &  &  &  &  & 111.65 & & 44.75 &  \\ 
\hline
\end{tabular}

  \caption{Results on ITC 2002 instances for 20 runs.}
  \label{tab:ITC2002_results}
\end{table}
\begin{table}
\centering
\footnotesize
\begin{tabular}{|c|r|rr|rr|r|r|rr|}\hline
 & \multicolumn{1}{c|}{N} & \multicolumn{ 2}{c|}{O} & \multicolumn{ 2}{c|}{P} & \multicolumn{1}{c|}{Q} & \multicolumn{1}{c|}{R} & \multicolumn{ 2}{c|}{Us} \\ 
Instance & \multicolumn{1}{c|}{Med} & \multicolumn{1}{c}{Avg} & \multicolumn{1}{c|}{Best} & \multicolumn{1}{c}{Avg} & \multicolumn{1}{c|}{Best} & \multicolumn{1}{c|}{Best} & \multicolumn{1}{c|}{Best} & \multicolumn{1}{c}{Avg} & \multicolumn{1}{c|}{Best} \\ 
\hline
    s1  & 1 & \textbf{0} & 0 & 0.8 & 0 & 3 & 0 & \textbf{0} & 0 \\ 
    s2  & 3 & \textbf{0} & 0 & 2 & 0 & 4 & 0 & 0.03 & 0 \\ 
    s3  & 1 & \textbf{0} & 0 & 1.3 & 0 & 6 & 0 & \textbf{0} & 0 \\ 
    s4  & 1 & \textbf{0} & 0 & 1 & 0 & 6 & 0 & 0.06 & 0 \\ 
    s5  & 0 & \textbf{0} & 0 & 0.2 & 0 & 0 & 0 & \textbf{0} & 0 \\ 
    m1  & 195 & 224.8 & 221 & 101.4 & 80 & 140 & 175 & \textbf{26.46} & 9 \\ 
    m2  & 184 & 150.6 & 147 & 116.9 & 105 & 130 & 197 & \textbf{25.86} & 15 \\ 
    m3  & 248 & 252 & 246 & 162.1 & 139 & 189 & 216 & \textbf{49.03} & 36 \\ 
    m4  & 164.5 & 167.8 & 165 & 108.8 & 88 & 112 & 149 & \textbf{23.83} & 12 \\ 
    m5  & 219.5 & 135.4 & 130 & 119.7 & 88 & 141 & 190 & \textbf{10.86} & 3 \\ 
    l1  & 851.5 & 552.4 & 529 & 834.1 & 730 & 876 & 912 & \textbf{259.8} & 208 \\     \hline

    Avg & & 134.82 & & 131.66 & & & &  \textbf{35.99} & \\     \hline
    l2  & --  & --  &  -- & --  & --  & --  & -- & 224.4 & 170 \\ 
    \hline
\end{tabular}
 
   \caption{Results on instances of the Metaheuristics Network for 30 runs.}
  \label{tab:socha_results}
\end{table}

The results of the experiments on the ITC 2002 instances are reported
in Table~\ref{tab:ITC2002_results}. Unfortunately, for all papers but
the one by \citeauthor{MNCR08} only best results are available from
the literature, therefore a fair comparison is not possible.
Nevertheless, it is clear from the table that the results of
\citeauthor{Kost04} are the overall bests. Regarding the comparison
with \citeauthor{MNCR08}, our solver clearly outperforms their one in
all 20 cases.

\subsubsection{Metaheuristics Network instances}

We finally move the the Metaheuristics Network instances, which have
been tackled by yet different authors. For these instances, our solver
greatly outperforms all the others on all the medium-size (m1--m5) and
on the first large instance (l1), while it is only marginally inferior
on the small ones. Instance l2 has not been tested recently in the
literature (because of the use of an incomplete copy of the dataset).

\section{Discussion, Conclusions, and Future Work}

We have presented a Simulated Annealing approach for a classical
well-studied timetabling problem, namely the PE-CTT problem.  The
comprehensive comparison with the literature shows that our solver is
able to outperform most of the previous approaches to the problem.
This result is obtained despite the fact that our solution is based on
a relatively simple single-step algorithm, whereas
most of the previous solvers use complex multi-step solution methods.  
In addition, the method proved to be quite robust w.r.t. the parameter values.

In our opinion, the key ingredients for these good results are the following.
First of all, the preprocessing and constraint reformulation steps
improve the efficacy of the local search. In particular, the
identification of the all-room events allows us to leave more space for
placing the other events.  Secondly, the room assignment procedure
based on attractiveness allows us to refrain from using the matching
algorithm, which is computationally expensive. Finally, the use of a
single-step procedure that takes into account the soft constraints from the
very beginning allowed us to save computational time later on during the search.

Only for the ITC 2002 instances the results are inferior to the best
ones reported in the literature. Unfortunately, the reliability of this
comparison is limited since it is based only on the best results found
by the other authors. 

For the future, we plan to extend our work in various directions:

\begin{enumerate}
\item Investigate on the use of different versions of Simulated
  Annealing, for example using different cooling schemes and
  acceptance criteria.
\item Improve our use of the tuning tools, mainly NOLHs and RACE, with
  the twofold objective of making them more effective and to
  automatise part of the experimental process.
\item Apply the same technique in different contexts, such as CB-CTT
  and other timetabling problems, in order to have a confirmation of
  its applicability.
\item Analyse the relevant features of the problem instances, in the
  spirit of \cite{KoSo04,SMLo11}, with the aim of obtaining an adaptive
  tuning. The idea would be to set the specific parameters based on
  the analysis of the specific instance, and the extraction of the
  values of specific features.
\end{enumerate}

\bibliographystyle{elsarticle-harv}
\bibliography{strings,timetabling,statistics}

\begin{thebibliography}{47}
\expandafter\ifx\csname natexlab\endcsname\relax\def\natexlab#1{#1}\fi
\expandafter\ifx\csname url\endcsname\relax
  \def\url#1{\texttt{#1}}\fi
\expandafter\ifx\csname urlprefix\endcsname\relax\def\urlprefix{URL }\fi

\bibitem[{Aarts and Lenstra(1997)}]{AaLe97}
Aarts, E., Lenstra, J.~K., 1997. Local Search in Combinatorial Optimization.
  John Wiley \& Sons, Chichester.

\bibitem[{Abdullah et~al.(2007)Abdullah, Burke, and McCollum}]{AbBM07b}
Abdullah, S., Burke, E.~K., McCollum, B., 2007. {A hybrid evolutionary approach
  to the university course timetabling problem}. In: Burke, E., Trick, M.~E.
  (Eds.), 2007 IEEE Congress on Evolutionary Computation. Vol. 3616.
  Springer-Verlag, pp. 1764--1768.

\bibitem[{Bellio et~al.(2011)Bellio, Di~Gaspero, and Schaerf}]{BeDS11}
Bellio, R., Di~Gaspero, L., Schaerf, A., 2011. Design and statistical analysis
  of a hybrid local search algorithm for course timetabling. Journal of
  Scheduling, 1--13.
\newline\urlprefix\url{http://dx.doi.org/10.1007/s10951-011-0224-2}

\bibitem[{Birattari(2005)}]{Bira05}
Birattari, M., April 2005. The RACE package.
\newline\urlprefix\url{http://cran.r-project.org/web/packages/race/}

\bibitem[{Burke et~al.(2003)Burke, Bykov, Newall, and Petrovi\'c}]{BBNP03}
Burke, E., Bykov, Y., Newall, J., Petrovi\'c, S., 2003. A time-predefined
  approach to course timetabling. Yugoslav Journal of Operations Research
  13~(2), 139--151.

\bibitem[{Burke et~al.(2010)Burke, Mare{\v c}ek, Parkes, and Rudov\'a}]{BMPR10}
Burke, E., Mare{\v c}ek, J., Parkes, A., Rudov\'a, H., 2010. A supernodal
  formulation of vertex colouring with applications in course timetabling.
  Annals of Operations Research 179~(1), 105--130.

\bibitem[{Burke and Petrovic(2002)}]{BuPe02}
Burke, E.~K., Petrovic, S., 2002. Recent research directions in automated
  timetabling. European Journal of Operational Research 140~(2), 266--280.

\bibitem[{Cambazard et~al.(2010)Cambazard, Hebrard, O'Sullivan, and
  Papadopoulos}]{CHOP10}
Cambazard, H., Hebrard, E., O'Sullivan, B., Papadopoulos, A., 2010. Local
  search and constraint programming for the post enrolment-based course
  timetabling problem. Annals of Operations Research, 1--25.
\newline\urlprefix\url{http://dx.doi.org/10.1007/s10479-010-0737-7}

\bibitem[{Chiarandini et~al.(2006)Chiarandini, Birattari, Socha, and
  Rossi-Doria}]{CBSR06}
Chiarandini, M., Birattari, M., Socha, K., Rossi-Doria, O., 2006. An effective
  hybrid approach for the university course timetabling problem. Journal of
  Scheduling 9~(5), 403--432.

\bibitem[{Chiarandini et~al.(2008)Chiarandini, Fawcett, and Hoos}]{ChFH08}
Chiarandini, M., Fawcett, C., Hoos, H., 2008. A modular multiphase heuristic
  solver for post enrolment course timetabling. In: Burke, E., Gendreau, M.
  (Eds.), Proc.\ of the 7th Int.\ Conf.\ on the Practice and Theory of
  Automated Timetabling (PATAT-2008). pp. 1--6.

\bibitem[{Cioppa and Lucas(2007)}]{CiLu07}
Cioppa, T.~M., Lucas, T.~W., 2007. Efficient nearly orthogonal and
  space-filling latin hypercubes. Technometrics 49~(1), 45--55.

\bibitem[{de~Werra(1985)}]{Werr85}
de~Werra, D., 1985. An introduction to timetabling. European Journal of
  Operational Research 19, 151--162.

\bibitem[{Di~Gaspero and Schaerf(2003)}]{DiSc03b}
Di~Gaspero, L., Schaerf, A., 2003. \textsc{EasyLocal++}: An object-oriented
  framework for flexible design of local search algorithms. Software---Practice
  and Experience 33~(8), 733--765.

\bibitem[{Di~Gaspero and Schaerf(2006)}]{DiSc06}
Di~Gaspero, L., Schaerf, A., 2006. Neighborhood portfolio approach for local
  search applied to timetabling problems. Journal of Mathematical Modeling and
  Algorithms 5~(1), 65--89.

\bibitem[{Eglese(1990)}]{Egle90}
Eglese, R.~W., 1990. Simulated annealing: a tool for operational research.
  European Journal of Operational Research 46, 271--281.

\bibitem[{Gotlieb(1963)}]{Gotl63}
Gotlieb, C.~C., 1963. The construction of class-teacher timetables. In:
  Popplewell, C.~M. (Ed.), {IFIP} congress 62. North-Holland, pp. 73--77.

\bibitem[{Hao and Benlic(2011)}]{HaBe11}
Hao, J.-K., Benlic, U., 2011. Lower bounds for the {ITC-2007} curriculum-based
  course timetabling problem. European Journal of Operational Research.
\newline\urlprefix\url{http://dx.doi.org/10.1016/j.ejor.2011.02.019}

\bibitem[{Hertz(1991)}]{Hert91}
Hertz, A., 1991. Tabu search for large scale timetabling problems. European
  Journal of Operational Research 54, 39--47.

\bibitem[{Hertz(1992)}]{Hert92}
Hertz, A., 1992. Finding a feasible course schedule using tabu search. Discrete
  Applied Mathematics 35~(3), 255--270.

\bibitem[{Hoos and St\"utzle(2005)}]{HoSt05}
Hoos, H.~H., St\"utzle, T., 2005. Stochastic Local Search -- Foundations and
  Applications. Morgan Kaufmann Publishers, San Francisco, CA (USA).

\bibitem[{Hutter et~al.(2009)Hutter, Hoos, Leyton-Brown, and
  St\"utzle}]{HHLS09}
Hutter, F., Hoos, H., Leyton-Brown, K., St\"utzle, T., 2009. Param{ILS}: An
  automatic algorithm configuration framework. Journal of Artificial
  Intelligence Research 36, 267--306.

\bibitem[{Johnson(2002)}]{John96}
Johnson, D.~S., 2002. A theoretician's guide to the experimental analysis of
  algorithms. In: Goldwasser, M.~H., Johnson, D.~S., McGeoch, C.~C. (Eds.),
  Data Structures, Near Neighbor Searches, and Methodology: Fifth and Sixth
  DIMACS Implementation Challenges. American Mathematical Society, pp.
  215--250.
\newline\urlprefix\url{http://www.research.att.com/~dsj/papers.html}

\bibitem[{Kirkpatrick et~al.(1983)Kirkpatrick, Gelatt, and Vecchi}]{KiGV83}
Kirkpatrick, S., Gelatt, Jr, C.~D., Vecchi, M.~P., 1983. Optimization by
  simulated annealing. Science 220, 671--680.

\bibitem[{Kostuch(2005)}]{Kost04}
Kostuch, P., 2005. The university course timetabling problem with a three-phase
  approach. In: Burke, E., Trick, M. (Eds.), Proc.\ of the 5th Int.\ Conf.\ on
  the Practice and Theory of Automated Timetabling (PATAT-2004), selected
  papers. Vol. 3616 of Lecture Notes in Computer Science. Springer-Verlag,
  Berlin-Heidelberg, pp. 109--125.

\bibitem[{Kostuch and Socha(2004)}]{KoSo04}
Kostuch, P., Socha, K., 2004. Hardness prediction for the university course
  timetabling problem. In: Gottlieb, J., Raidl, G.~R. (Eds.), Evolutionary
  Computation in Combinatorial Optimization. Vol. 3004 of Lecture Notes in
  Computer Science. Springer Berlin / Heidelberg, pp. 135--144.

\bibitem[{Lach and L\"ubbecke(2011)}]{LaLu10}
Lach, G., L\"ubbecke, M., 2011. Curriculum based course timetabling: Optimal
  solutions to the udine benchmark instances. Annals of Operations Research,
  1--18.
\newline\urlprefix\url{http://dx.doi.org/10.1007/s10479-010-0700-7}

\bibitem[{Landa-Silva and Obit(2008)}]{LaOb08}
Landa-Silva, D., Obit, J.~H., September 2008. {Great deluge with non-linear
  decay rate for solving course timetabling problems}. In: 2008 4th
  International IEEE Conference Intelligent Systems. pp. 8--18.

\bibitem[{Lewis(2008)}]{Lewi08}
Lewis, R., 2008. A survey of metaheuristic-based techniques for university
  timetabling problems. OR Spectrum 30~(1), 167--190.

\bibitem[{Lewis(2010)}]{Lewi10}
Lewis, R., 2010. A time-dependent metaheuristic algorithm for post
  enrolment-based course timetabling. Annals of Operations Research, 1--17.
\newline\urlprefix\url{http://dx.doi.org/10.1007/s10479-010-0696-z}

\bibitem[{Lewis and Paechter(2005)}]{LePa05}
Lewis, R., Paechter, B., 2005. Application of the grouping genetic algorithm to
  university course timetabling. In: Raidl, G., Gottlieb, J. (Eds.), Fifth
  International Conference on Evolutionary Computation in Combinatorial
  Optimisation (EvoCop 2005). No. 3448 in LNCS. Springer, pp. 144--153.

\bibitem[{Lewis and Paechter(2007)}]{LePa07}
Lewis, R., Paechter, B., 2007. Finding feasible timetables using group-based
  operators. IEEE Transactions on Evolutionary Computation 11~(3), 397--413.

\bibitem[{Lewis et~al.(2007)Lewis, Paechter, and McCollum}]{LePM07}
Lewis, R., Paechter, B., McCollum, B., 2007. Post enrolment based course
  timetabling: A description of the problem model used for track two of the
  second international timetabling competition. Tech. rep., Cardiff University,
  Wales, UK.

\bibitem[{Liu et~al.(2011)Liu, Zhang, and Chin}]{LiZC11}
Liu, Y., Zhang, D., Chin, F., 2011. A clique-based algorithm for constructing
  feasible timetables. Optimization Methods and Software 26~(2), 281--294.

\bibitem[{L\"u and Hao(2009)}]{LuHa09}
L\"u, Z., Hao, J.-K., 2009. Adaptive tabu search for course timetabling.
  European Journal of Operational Research 200~(1), 235 -- 244.

\bibitem[{Mayer et~al.(2008)Mayer, Nothegger, Chwatal, and Raidl}]{MNCR08}
Mayer, A., Nothegger, C., Chwatal, A., Raidl, G., 2008. Solving the post
  enrolment course timetabling problem by ant colony optimization. In: Burke,
  E., Gendreau, M. (Eds.), Proc.\ of the 7th Int.\ Conf.\ on the Practice and
  Theory of Automated Timetabling (PATAT-2008). pp. 1--13.

\bibitem[{McCollum et~al.(2010)McCollum, Schaerf, Paechter, McMullan, Lewis,
  Parkes, Di~Gaspero, Qu, , and Burke}]{MSPM10}
McCollum, B., Schaerf, A., Paechter, B., McMullan, P., Lewis, R., Parkes,
  A.~J., Di~Gaspero, L., Qu, R., , Burke, E.~K., 2010. Setting the research
  agenda in automated timetabling: The second international timetabling
  competition. {INFORMS} Journal on Computing 22~(1), 120--130.

\bibitem[{McMullan(2007)}]{McMu07}
McMullan, P., 2007. {An Extended Implementation of the Great Deluge Algorithm
  for Course Timetabling}. Computational Science--ICCS 2007, 538 -- 545.

\bibitem[{M\"uller(2009)}]{Mull09}
M\"uller, T., 2009. {ITC2007} solver description: a hybrid approach. Annals of
  Operations Research 172~(1), 429--446.

\bibitem[{Murray et~al.(2007)Murray, M{\"u}ller, and Rudov{\'a}}]{MuMR07}
Murray, K.~S., M{\"u}ller, T., Rudov{\'a}, H., 2007. Modeling and solution of a
  complex university course timetabling problem. In: Proc.\ of the 6th Int.\
  Conf.\ on the Practice and Theory of Automated Timetabling (PATAT-2006),
  selected papers. pp. 189--209.

\bibitem[{Rossi-Doria et~al.(2003)}]{RSBC03}
Rossi-Doria, O., et~al., 2003. A comparison of the performance of different
  metaherustic on the timetabling problem. In: Burke, E., De~Causmaecker, P.
  (Eds.), Proc.\ of the 4th Int.\ Conf.\ on the Practice and Theory of
  Automated Timetabling (PATAT-2002), selected papers. Vol. 2740 of Lecture
  Notes in Computer Science. Springer-Verlag, Berlin-Heidelberg, pp. 329--351.

\bibitem[{Sanchez(2005)}]{Sanc05}
Sanchez, S.~M., 2005. {NOLH} designs spreeadsheet. Visited on April 6, 2011.
\newline\urlprefix\url{http://diana.cs.nps.navy.mil/SeedLab/}

\bibitem[{Schaerf(1999)}]{Scha99}
Schaerf, A., 1999. A survey of automated timetabling. Artificial Intelligence
  Review 13~(2), 87--127.

\bibitem[{Smith-Miles and Lopes(2011)}]{SMLo11}
Smith-Miles, K.~A., Lopes, L., 2011. Generalising algorithm performance in
  instance space: A timetabling case study. In: Proceedings of Learning and
  Intelligent Optimization (LION 2011). pp. 1--15.

\bibitem[{Socha et~al.(2002)Socha, Knowles, and Sampels}]{SoKS02}
Socha, K., Knowles, J., Sampels, M., 2002. A {MAX-MIN} ant system for the
  university timetabling problem. In: Proc.\ of the 3rd International Workshop
  on Ant Algorithms (ANTS 2002). Vol. 2463 of Lecture Notes in Computer
  Science. Springer-Verlag, Berlin, Germany, pp. 1--13.

\bibitem[{Tuga et~al.(2007)Tuga, Berretta, and Mendes}]{TuBM07}
Tuga, M., Berretta, R., Mendes, A., 2007. A hybrid simulated annealing with
  kempe chain neighborhood for the university timetabling problem. In: 6th
  {IEEE/ACIS} International Conference on Computer and Information Science
  (ICIS 2007). pp. 400 -- 405.

\bibitem[{Turabieh et~al.(2009)Turabieh, Abdullah, and McCollum}]{TuSM09}
Turabieh, H., Abdullah, S., McCollum, B., 2009. {Electromagnetism-like
  Mechanism with Force Decay Rate Great Deluge for the Course Timetabling
  Problem}. Lecture Notes In Artificial Intelligence; Vol. 5589, 497.

\bibitem[{van~den Broek and Hurkens(2010)}]{BrHu10}
van~den Broek, J. J.~J., Hurkens, C. A.~J., 2010. An {IP}-based heuristic for
  the post enrolment course timetabling problem of the {ITC2007}. Annals of
  Operations Research, 1--16.
\newline\urlprefix\url{http://dx.doi.org/10.1007/s10479-010-0708-z}

\end{thebibliography}

\end{document}